# Determination of a negative curvature of the Universe from the magnitude-redshift relation of high redshift supernovae

Mikolaj „Mik" Sawicki

*John A. Logan College*
*Carterville, IL 62918*

email: mikolaj.sawicki@jal.cc.il.us

*Abstract:*

*Recent observations of high redshift Supernovae at lower than expected value of the Hubble constant, widely interpreted as an evidence for accelerating expansion of the Universe, could alternatively be explained assuming a hyperbolic Universe with a negative spatial curvature of the order of k = - 1/(2100 Mpc).*

Calan/Tololo Supernova Survey of local Type Ia supernovae (these with redshifts smaller than 0.1 and distances mainly between 20 and 300 Mpc) led to a determination of the Hubble constant as $H_0 = (72 \pm 8)$ km/s/Mpc [1].

Two independent groups searching for distant supernovae, the Supernova Cosmology Project [2,3,4] (at redshifts between 0.18 and 0.83), and the High-Z Supernova Team [5,6,7,8] obtained equivalent and unexpected results indicating that "high redshift" supernovae at $z \sim 0.5$ appear to be approximately 0.2 magnitude fainter than expected on the basis of the above quoted local value of the Hubble constant. This discovery is widely interpreted as evidence that these supernovae are more distant than expected, thus yielding a reduced value of the Hubble constant for these distant objects, whose look-back time is some 5 billion years. This gives rise to the postulate of the accelerating expansion of the Universe; since in the past the far-away regions of the Universe expanded at a lower rate than our local vicinity today, the expansion must have accelerated [9].

Specifically, using the value $H_0 = 72$ km/s/Mpc, a $z = 0.5$ supernova was expected to be at a distance of 1600 Mpc. The observed change of its brightness by 0.2 mag corresponds to a reduction of the observed intensity by a factor of 1.2, and hence could be attributed to an increase of its distance by a factor of $\sqrt{1.2} = 1.1$. This in turn implies a reduction of the Hubble constant for this object by about 10%.

This is however based on an assumption that the Universe is flat and hence the area of a sphere of radius $r$ is simply $4\pi r^2$. For a hyperbolic Universe with a negative curvature $k = -1/R$, where $k < 0$ and $R > 0$, the surface area of a sphere of radius $r$ is

$$A(r) = \frac{4\pi}{k^2} \sinh^2(kr) \qquad (1)$$

For small distances, $r << R$, Eq.(1) reduces to

$$A(r) = 4\pi r^2 \left(1 + \frac{1}{3}\frac{r^2}{R^2}\right) \qquad (2)$$

The area of a sphere in the hyperbolic space is therefore by a factor $\chi$ larger than the area of a sphere in a flat space, where

$$\chi = \left(\frac{\sinh(kr)}{kr}\right)^2 \qquad (3)$$

and the increase in the apparent magnitude of the supernova is given by

$$\Delta m = \frac{\log \chi}{\log(2.512)} \qquad (4)$$

To obtain $\Delta m = 0.2$ for $r = 1600$ Mpc, the curvature $k$ should have a value of $k = -4.7 \cdot 10^{-4}$ Mpc$^{-1}$, i.e. the parameter $R$ should be equal to $R = 2.1 \cdot 10^3$ Mpc.

In Figure 1 we show expected increase in the apparent magnitude as a function of the supernova distance, calculated from the Hubble law with the low redshift value of $H_0 = (72\pm8)$ km/s/Mpc. Extrapolation to larger distances yields the expected value of $\Delta m = 0.3$ for $r = 2000$ Mpc ($z = 0.687$), and $\Delta m = 0.67$ for $r = 3000$ Mpc ($z = 1.48$). We hope these predictions could be tested experimentally in a near future.

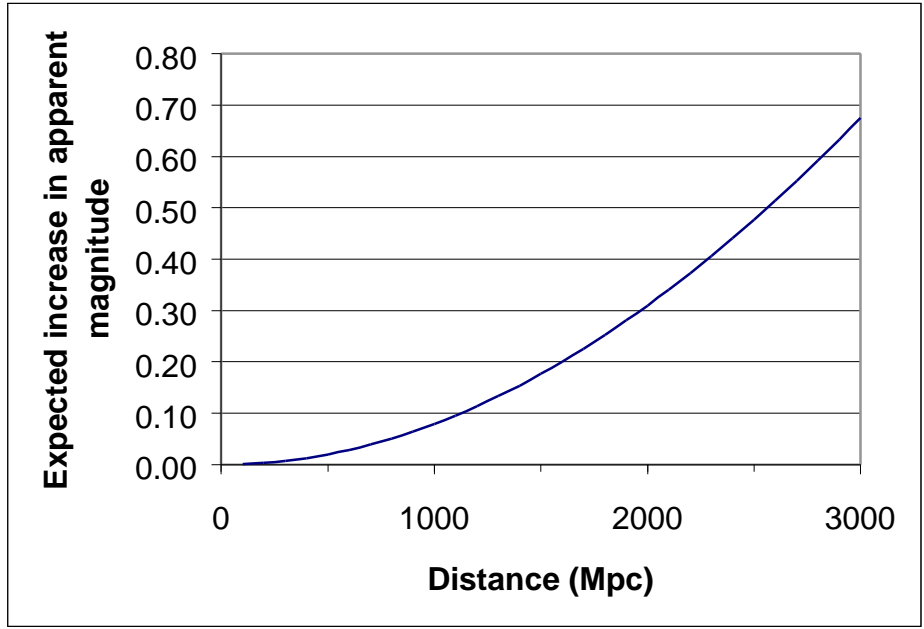

*Figure 1. Expected increase in apparent magnitude of a supernova as a function of the distance determined from a redshift using local value of the Hubble constant $H_0$ = 72 km/s/Mpc, assuming hyperbolic spatial curvature of the Universe.*

**Acknowledgements**

I wish to thank Jan Mycielski for a tutorial on surfaces in a hyperbolic space, and to acknowledge discussions with Josh Frieman and Keith Olive during the TASI 2002 in Boulder, Colorado.

**References**

1. M. Hamuy *et al*., Astron. Journ. **112**, 2438 (1966) [astro-ph/9609059].
2. S. Perlmutter *et al,* [Supernova Cosmology Project Collaboration], Astrophys. Journ. **483**, 565 (1997) [astro-ph/9608192].
3. S. Perlmutter *et al,* [Supernova Cosmology Project Collaboration], Nature **391**, 51 (1998) [astro-ph/9712212].
4. S. Perlmutter *et al,* [Supernova Cosmology Project Collaboration], Astrophys. Journ. **517**, 565 (1999) [astro-ph/9812133].
5. P.M. Garnavich *et al*. [Hi-Z Supernova Team Collaboration], Astrophys. Journ. **493**, L53 (1998) [astro-ph/9710123].
6. B.P. Schmidt *et al*. [Hi-Z Supernova Team Collaboration], Astrophys. Journ. **507**, 46 (1998) [astro-ph/9805200].
7. A.G. Riess *et al*. [Hi-Z Supernova Team Collaboration], Astron. Journ. **116**, 1009 (1998) [astro-ph/9805201].
8. P.M. Garnavich *et al*. [Hi-Z Supernova Team Collaboration], Astrophys. Journ. **509**, 74 (1998) [astro-ph/9806396].
9. Josh Frieman, TASI lectures, June 2002 (unpublished).